\documentclass[letter]{aa}  
\usepackage{graphicx}
\usepackage[varg]{txfonts}
\usepackage[utf8]{inputenc}
\usepackage[T1]{fontenc}
\usepackage[urlcolor = black, citecolor = blue, colorlinks = True]{hyperref}
\usepackage{float}

\begin{document}

\title{A slow bar in the lenticular barred galaxy NGC~4277\thanks{Based on observations made with ESO telescopes at the La Silla-Paranal Observatory under programme 094.B-0241.}}

\author{C. Buttitta\inst{1,}\thanks{\email{chiara.buttitta@phd.unipd.it}}
       \and E.~M. Corsini\inst{1,2} 
       \and V. Cuomo\inst{3}
       \and J.~A.~L. Aguerri\inst{4,5}
       \and L. Coccato\inst{6}
       \and L. Costantin\inst{7} 
       \and E. Dalla Bont\`a\inst{1,2} 
       \and \\ V.~P. Debattista\inst{8}
       \and E. Iodice\inst{9}
       \and J. M\'endez-Abreu\inst{4,5}
       \and L. Morelli\inst{3}
       \and A. Pizzella\inst{1,2} 
       }

\institute{
Dipartimento di Fisica e Astronomia ``G. Galilei'', Universit\`a di Padova, vicolo dell'Osservatorio 3, I-35122 Padova, Italy
\and
INAF-Osservatorio Astronomico di Padova, vicolo dell'Osservatorio 2, I-35122 Padova, Italy 
\and 
Instituto de Astronom\'{\i}a y Ciencias Planetarias, Universidad de Atacama, Avenida Copayapu 485, Copiap\'o, Chile
\and 
Departamento de Astrof\'{\i}sica, Universidad de La Laguna, Avenida Astrof\'{\i}sico Francisco S\'anchez s/n, 38206 , Tenerife, Spain
\and
Instituto de Astrof\'{\i}sica de Canarias, calle V\'{\i}a L\'actea s/n, 38205 La Laguna, Tenerife, Spain
\and
European Southern Observatory, Karl-Schwarzschild-Strasse 2, D-85748 Garching, Germany
\and
Centro de Astrobiolog\'{\i}a (CSIC-INTA), Ctra de Ajalvir km 4, Torrej\'on de Ardoz, 28850, Madrid, Spain
\and
Jeremiah Horrocks Institute, University of Central Lancashire, Preston, PR1 2HE  UK
\and
INAF-Osservatorio Astronomico di Capodimonte, via Moiariello 16, I-80131 Napoli, Italy
}

\date{}

\abstract
  {}
  {We characterised the properties of the bar hosted in lenticular galaxy NGC\,4277, which is located behind the Virgo cluster.}
  {We measured the bar length and strength from the surface photometry obtained from the broad-band imaging of the Sloan Digital Sky Survey and we derived the bar pattern speed from the stellar kinematics obtained from the integral-field spectroscopy performed with the Multi Unit Spectroscopic Explorer at the Very Large Telescope. We also estimated the co-rotation radius from the circular velocity, which we constrained by correcting the stellar streaming motions for asymmetric drift, and we finally derived the bar rotation rate.}
  {We found that NGC\,4277 hosts a short ($R_{\rm bar}=3.2^{+0.9}_{-0.6}$\,kpc), weak ($S_{\rm bar}=0.21 \pm 0.02$), and slow (${\cal{R}}=1.8^{+0.5}_ {-0.3}$) bar and its pattern speed ($\Omega_{\rm bar}=24.7\pm3.4$ km s$^{-1}$ kpc$^{-1}$) is amongst the best-constrained ones ever obtained with the Tremaine-Weinberg (TW) method with relative statistical errors of $\sim0.2$.}
  {NGC\,4277 is the first clear-cut case of a galaxy hosting a slow stellar bar (${\cal{R}}>1.4$ at more than a 1$\sigma$ confidence level) measured with the model-independent TW method. A possible interaction with the neighbour galaxy NGC\,4273 could have triggered the formation of such a slow bar and/or the bar could be slowed down due to the dynamical friction with a significant amount of dark matter within the bar region.}

\maketitle

\section{Introduction}
\label{sec:introduction}

Stellar bars are ubiquitous in nearby disc and irregular galaxies \citep[e.g.][]{Marinova2007, Aguerri2009, Buta2015}. The (triggered and spontaneous) formation, subsequent (fast and slow) evolution, and possible (abrupt and progressive) dissolution of a bar drive remarkable changes in the host galaxy on both small ($\sim100$ pc) and large ($\sim10$ kpc) spatial scales over both short ($\sim100$ Myr) and long ($\sim10$ Gyr) timescales. Indeed, the exchange of mass, energy, and angular momentum between the bar and the other components of the galaxy affects its morphology, orbital structure, mass distribution, star formation, central fuelling rate, and stellar population properties \citep{Kormendy2004, Athanassoula2013, Fragkoudi2016}. This metamorphosis of the galaxy is tightly coupled to the changes expected in the properties of the bar. Once born, bars become longer and stronger as well as slow down on timescales, which also depend on the dynamical friction generated by the dark matter (DM) halo in the bar region \citep[e.g.][]{Debattista1998, Athanassoula2013, Petersen2019}

Measuring the length $R_{\rm bar}$, strength $S_{\rm bar}$, and pattern speed $\Omega_{\rm bar}$ of bars is therefore vital for unveiling the structure of barred galaxies (see \citealt{Athanassoula2013} and \citealt{Sellwood2014}, for reviews). We note that $R_{\rm bar}$ corresponds to the radial extension of the stellar orbits that support the bar, $S_{\rm bar}$ parameterises the bar's contribution to the galaxy's non-axisymmetric potential, and $\Omega_{\rm bar}$ is the bar pattern speed. A further parameter, the rotation rate, $\mathcal{R}$ is defined as the ratio between the co-rotation radius and the bar length and it classifies bars into `fast and long' ($1\leq \mathcal{R} \leq 1.4$) and `slow and short' bars ($\mathcal{R}>1.4$)  \citep{Athanassoula1992, Debattista2000}. Fast and long bars can form spontaneously in unstable and nearly isolated stellar discs \citep{Athanassoula2013}, while the formation of slow and short bars can be induced by the tidal interaction with a neighbour galaxy \citep{Martinez-Valpuesta2017}. Moreover, the bar can be braked by the DM halo, and therefore $\mathcal{R}$ could be a good proxy for the content and distribution of DM in the bar region \citep[e.g.][]{Debattista1998, Athanassoula2013, Petersen2019}.

While $R_{\rm bar}$ and $S_{\rm bar}$ can be determined through the analysis of the surface brightness distribution (see \citealt{Aguerri2009}, and references therein), $\Omega_{\rm bar}$ is a kinematic parameter which requires both photometric and kinematic data. In the last decades, several indirect methods have been proposed to recover $\Omega_{\rm bar}$. They are based on the identification of rings with resonances \citep{Perez2012}, the study of the shape of dust lanes \citep{Athanassoula1992}, the location of shock-induced star-forming regions \citep{Puerari1997}, the comparison of dynamical models of gas \citep{Weiner2001} or N-body simulations \citep{Rautiainen2008} with the observed distribution of gas and stars, the analysis of the phase shift between the bar density perturbation and gravitational potential \citep{Zhang2007}, and the assessment of the phase change of the gas flow across co-rotation \citep{Font2011}. All of these methods are model dependent and suffer some limitations. For example, the correct identification of resonances and shock-induced star-forming regions is not a straightforward task; the dust lanes across the bar are often very subtle and quite complex features, and gas dynamical models and N-body simulations lead to non-unique solutions when compared to the actual morphology of a galaxy.
The only model-independent technique that can recover $\Omega_{\rm bar}$ is the \cite{TW1984} method (hereafter TW method), which is based on the assumptions that the bar rotates with a well-defined pattern speed in a flat disc and that the tracer satisfies the continuity equation resulting in $\langle V \rangle=\langle X \rangle \sin{(i)}\,\Omega_{\rm bar}$, where $i$ is the disc inclination and $\langle V \rangle$ and $\langle X \rangle $ are the mass-weighted mean line-of-sight (LOS) velocities and positions of the tracer measured along different apertures parallel to the disc major axis. 

To date no `genuine' slow bar ($\mathcal{R}>1.4$ at more than a 1$\sigma$ confidence level) have been measured using the TW method on stars (see \citealt{Corsini2011} and \citealt{Cuomo2020}, for the full list of objects), suggesting that bar formation was not tidally induced by close interactions and implying a low DM contribution in the bar region. Only a few slow bars have been found by applying the TW method to a gaseous tracer, such as the neutral \citep{Banerjee2013, Patra2019} or ionised hydrogen one \citep{Bureau1999, Chemin2009}, with no guarantee that the continuity equation holds in the presence of gas phase changes and ongoing star formation.
On the contrary, a large number of bars have been found to be slow through model-dependent methods. However, either we do not have a  reliable estimate of the uncertainty on $\mathcal{R}$ or the uncertainty is so large ($\Delta \mathcal{R}/ \mathcal{R}\geq0.5$) that these bars are also consistent with being `fast' \citep{Rautiainen2008, ButaZhang2009, Font2014}. Here, we report the case of NGC\,4277 as the first clear-cut example of a galaxy hosting a slow stellar bar, from the direct measurement of its pattern speed. 

\section{NGC\,4277}
\label{sec:n4277}

The lenticular barred galaxy NGC\,4277 is an ideal target for the application of the TW method to accurately measure $\Omega_{\rm bar}$. It has an intermediate inclination, its bar is oriented at an intermediate angle between the major and minor axes of the disc, and it shows no evidence of spiral arms or patchy dust (Fig.~\ref{fig:pointings}). NGC\,4277 is classified as SBa by \cite{Binggeli1985}, SAB(rs)0/a by \citet[][ {hereafter RC3}]{RC3}, SB0$^0$ by \citet{Baillard2011}, and SAB(r\underline{s})0$^+$ by \citet{Buta2015}. It has an apparent magnitude $B_{T}=13.38$ mag (RC3), which corresponds to a total absolute corrected magnitude $M^{0}_{B_{\rm T}} = -19.27$ mag, obtained adopting a distance $D = 33.9$ Mpc from the systemic velocity with respect to the cosmic microwave background reference frame $V_{\rm CMB} = 2542 \pm 48$\ km\,s$^{-1}$ \citep{Fixsen1996} and assuming $H_0=75$ km s$^{-1}$ Mpc$^{-1}$. The stellar mass is $M_\star = 8 \cdot 10^9$ M$_{\odot}$ with a lower limit for the \ion{H}{i} mass of $M_{\rm HI} = 7 \cdot 10^8$ M$_{\odot}$ for the adopted distance \citep{vanDriel2016}. The galaxy possibly forms an interacting pair with NGC\,4273 \citep{Kim2014}. The latter lies at a projected distance of 1\farcm9 and it is located at a distance $D = 36.3$ Mpc. The two galaxies are both classified as possible members of the Virgo cluster \citep{Kim2014}.

\section{Observations and data reduction}
\label{sec:observations}

\begin{figure}[t]
    \centering
    \includegraphics[scale=0.33]{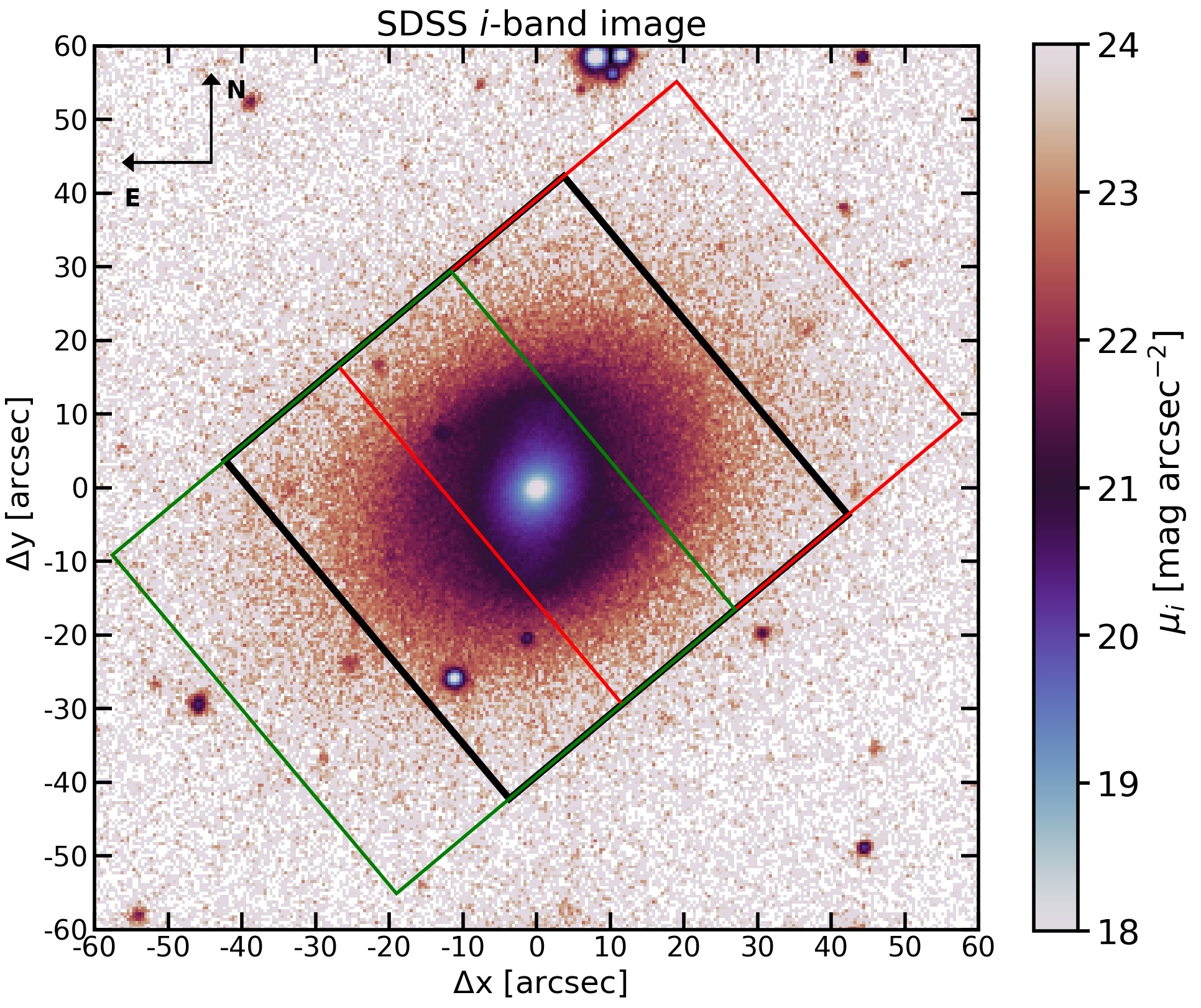}
    \caption{SDSS $i$-band image of NGC\,4277. The three squares mark the MUSE central (solid black lines) and offset (red and green lines) pointings. They cover a total FOV of $1\farcm7 \times 1\farcm0$.}
    \label{fig:pointings}
\end{figure}

We carried out the spectroscopic observations of NGC\,4277 in service mode on 20 March 2015 (Prog. Id. 094.B-0241(A); P.I.: E.M. Corsini) with the Multi Unit Spectroscopic Explorer \citep[MUSE,][]{Bacon2010} of the European Southern Observatory. We configured MUSE in wide field mode to ensure a nominal field of view (FOV) of $1'\times1'$ with a spatial sampling of $0\farcs2$ pixel$^{-1}$ and to cover the wavelength range of 4800\,--\,9300 \AA\ with a spectral sampling of 1.25 \AA\ pixel$^{-1}$ and an average nominal spectral resolution of $\mathrm{FWHM} = 2.51$ \AA. We split the observations into two observing blocks (OBs) to map the entire galaxy along its photometric major axis. We organised each OB to perform four pointings. The first pointing was centred on the nucleus, the second one on a blank sky region at a few arcmins from the galaxy, and the third and fourth ones were an eastward and westward offset along the galaxy's major axis taken at a distance of $20''$ from the  galactic nucleus, respectively (Fig.~\ref{fig:pointings}). The mean value of the seeing during the observations was $\mathrm{FWHM}\sim1\farcs1$. We performed the standard data reduction with the MUSE pipeline \citep{Weilbacher2016} under the {\sc esoreflex} environment \citep{Freudling2013} to obtain the combined datacube of NGC\,4277. Finally, we subtracted the residual sky signal following \citet{Cuomo2019}. In addition to spectroscopic data, we retrieved the flux-calibrated $i$-band image of NGC\,4277 from the science archive of Sloan Digital Sky Survey (SDSS) Data Release 14 \citep[]{SDSSDR14} and subtracted the residual sky level as was done in \citet{Morelli2016}. 

\section{Data analysis and results}
\label{sec:analysis}

\begin{figure*}
\centering
\includegraphics[scale=0.39]{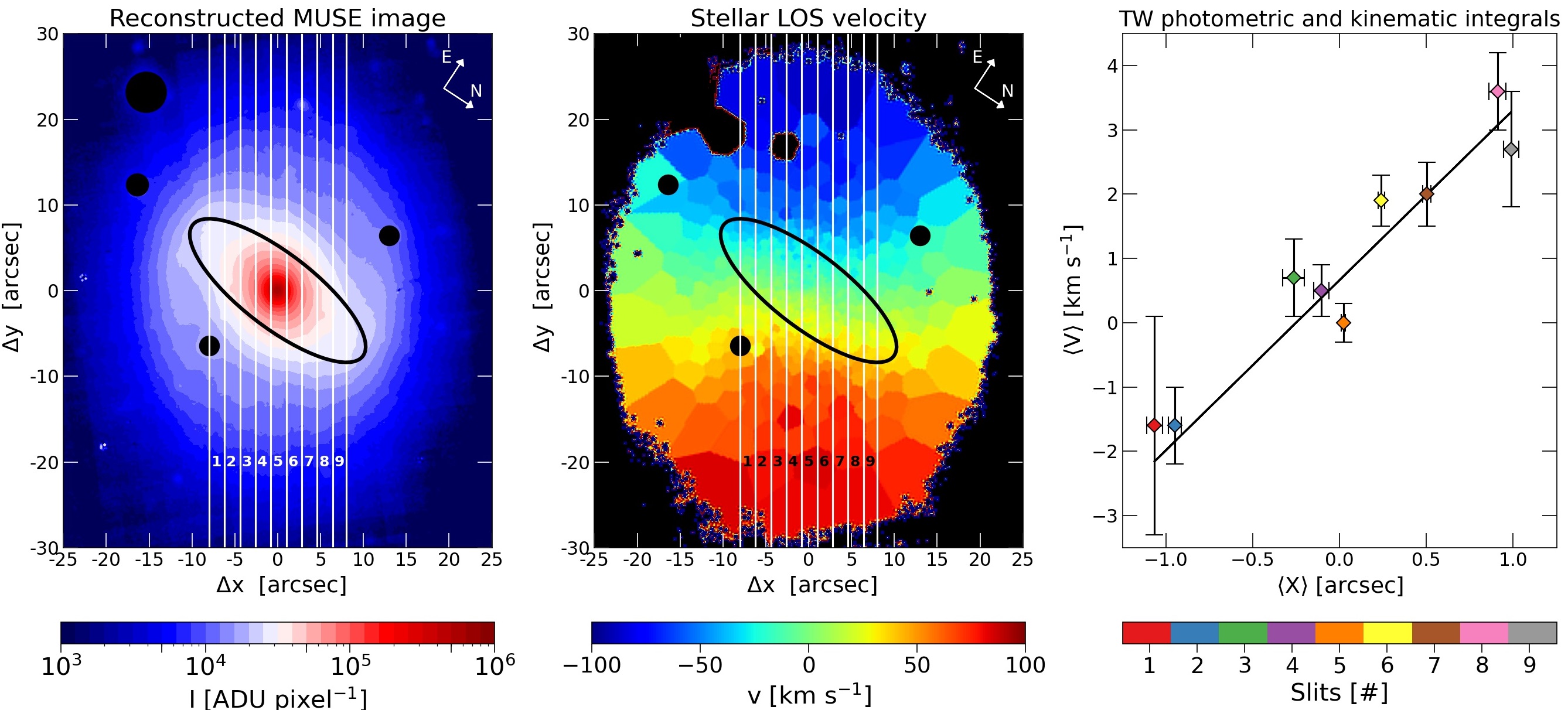}
\caption{MUSE data of NGC\,4277. {\itshape Left panel}: MUSE reconstructed image with pseudo-slits (white lines) and a bar isophote (black ellipse). The FOV is 50\,$\times$\,60 arcmin$^2$ and the disc major axis is parallel to the vertical axis. {\itshape Central panel}: Mean stellar LOS velocity map of NGC~4277 with a bar isophote (black ellipse). The FOV is 50\,$\times$\,60 arcmin$^2$ and the disc major axis is parallel to the vertical axis. {\itshape Right panel}: Kinematic integrals $\langle V\rangle$ as a function of photometric integrals $\langle X\rangle$. The black solid line represents the best fit to the data.}
\label{fig:tw}
\end{figure*}

\subsection{Bar length and strength}
\label{sec:photometry}

We performed the isophotal analysis of the $i$-band image of NGC\,4277 using the {\sc ellipse} task in {\sc iraf} \citep{Jedrzejewski1987}. We fitted the galaxy isophotes with ellipses fixing the centre coordinates after checking they do not vary within the uncertainties. The radial profiles of the azimuthally  averaged surface brightness, $\mu$, position angle, PA, and ellipticity, $\epsilon$, are shown in Fig.~\ref{fig:photometry} (left panels) as a function of the semi-major axis of the ellipses, $a$. These profiles show the typical trends observed in barred galaxies \citep[e.g.][]{Aguerri2000}: $\epsilon$ exhibits a local maximum and the PA is nearly constant in the bar region, while $\epsilon$ and PA are both constant in the disc region. We estimated the mean geometric parameters of the disc  ($\langle\epsilon\rangle = 0.242 \pm 0.002$, $\langle {\rm PA} \rangle = 123\fdg27 \pm 0\fdg32$) in the radial range $28''\leq a \leq 48''$ following the prescriptions by \citet{Cuomo2019}. We assumed that the disc is infinitesimally thin to estimate the galaxy inclination $i = 40\fdg7 \pm 0\fdg2$. 
We adopted the disc geometric parameters to deproject the galaxy image and then we fitted ellipses to the resulting isophotes. We measured the bar length $R_{\rm PA} = 18\farcs2 \pm 0\farcs4$ as the radius where the PA changes by $10\degr$ from the PA of the ellipse with the maximum $\epsilon$, as in \citet{Aguerri2009}. 

We then analysed the deprojected $i$-band image of NGC\,4277 to carry out the Fourier analysis of the azimuthal surface-brightness distribution of the galaxy. We derived the radial profiles of the amplitudes of the $m=0,1,\dots,6$ Fourier components and of the phase angle $\phi_2$ of the $m=2$ Fourier component as was done in \citet{Aguerri2000}. They displayed the same behaviour as measured in other galaxies hosting an elongated bi-symmetric bar \citep[e.g.][]{Ohta1990}: the amplitudes of the even Fourier components are larger than those of the odd ones, with the $m=2$ component having a prominent peak and constant phase angle $\phi_2$ in the bar region. The radial profiles of the relative amplitudes of the $m=2,4,6$ Fourier components are shown in Fig.~\ref{fig:photometry} (right panels). We measured the bar length $R_{\rm bar/ibar} = 19\farcs0\pm2\farcs2$ from the bar-interbar intensity ratio derived from the amplitudes of the Fourier components  \citep{Aguerri2000} and $R_{\phi_2} = 15\farcs8\pm4\farcs5$ from the analysis of the phase angle $\phi_2$ \citep{Debattista2002}. We also estimated the bar strength $S_{\rm Fourier} = 0.196\pm0.004$ from the mean value of the $I_2/I_0$ ratio over the bar extension as in \cite{Athanassoula2002}.

We derived the structural parameters of NGC\,4277 by performing a photometric decomposition of the $i$-band image with the {\sc gasp2d} algorithm \citep{Mendez-Abreu2008, MendezAbreu2017}. We adopted a S\'ersic bulge \citep{Sersic1968}, a double-exponential disc \citep{MendezAbreu2017}, and a Ferrers bar (\citealt{Aguerri2009}) to model the galaxy surface-brightness distribution. The best-fitting values of the structural parameters, including the length and axial ratio of the bar ($R_{\rm Ferrers}=25\farcs0\pm0\farcs1$, $q_{\rm Ferrers} = 0.341\pm0.001$), were constrained by performing a $\chi^2$ minimisation, accounting for the photon noise, read-out noise, and point spread function of the image. The best-fitting values together with their errors, which we estimated by analysing a sample of images of mock galaxies built with Monte Carlo simulations, are reported in Table~\ref{tab:gasp2d}.  The results of the photometric decomposition of NGC\,4277 obtained with {\sc gasp2d} are shown in Fig.~\ref{fig:gasp2d}. 

We adopted the mean of $R_{\rm PA}$, $R_{\rm bar/ibar}$, $R_{\phi_2}$, and $R_{\rm Ferrers}$ as the length $R_{\rm bar}$ of the bar and we calculated its $\pm1\sigma$ error as the difference with respect to highest and lowest measured value. This gives $R_{\rm bar}= 19\farcs5^{+5\farcs5}_{-3\farcs7}$, which corresponds to $3.2^{+0.9}_{-0.6}$ kpc at the assumed distance. We compared this value with the typical bar length of SB0 galaxies measured by \citet{Aguerri2009} and we conclude that NGC\,4277 hosts a short bar. We derived the bar strength $S_\epsilon = 0.230 \pm 0.003$ from the ellipticity at $R_{\rm bar}$ measured on the deprojected galaxy image following \citet{Aguerri2009}. We took the mean value of $S_{\rm Fourier}$ and $S_\epsilon$ and their semi-difference as the strength $S_{\rm bar}$ of the bar and its error, respectively. This gives $S_{\rm bar} = 0.21 \pm 0.02$, which means that the bar of NGC\,4277 is weak according to the classification of \citet{Cuomo2019b}.

\subsection{Stellar kinematics and circular velocity}
\label{sec:kinematics}

We measured the stellar and ionised-gas kinematics of NGC\,4277 from the sky-cleaned MUSE datacube using the {\sc ppxf} \citep{Cappellari2004} and {\sc gandalf} \citep{Sarzi2006} algorithms. We spatially binned the datacube spaxels with the adaptive algorithm of \cite{Cappellari2003} based on Voronoi tessellation to obtain a target $S/N = 40$ per bin. We used the ELODIE library ($\sigma_{\rm instr}=13$ km s$^{-1}$, \citealt{Prugniel2001}) in the wavelength range 4800\,--\,5600 \AA\ centred on the \ion{Mg}{i}$\lambda\lambda5167,5173,5184$ absorption-line triplet and including the [\ion{O}{iii}]$\lambda\lambda4959,5007$ and [\ion{N}{i}]$\lambda\lambda5198,5200$ emission-line doublets. 
The maps of the LOS velocity $v$ and velocity dispersion $\sigma$ of the stellar component are shown in Fig.~\ref{fig:kinematics}. We estimated the errors on $v$ and $\sigma$ from formal errors of the {\sc ppxf} best fit as was done in \cite{Corsini2018}; they range between 1 and 18 km s$^{-1}$. We calculated the residual noise $rN$ as the standard deviation of the difference between the galaxy and the best-fitting stellar spectrum. Finally, we simultaneously fitted the ionised-gas emission lines with Gaussian functions. We did not detect any emission line, except for a few isolated spatial bins in the disc region. In these bins, the signal-to-residual noise of the emission lines is $S/rN \ga 3$.

We derived the circular speed $V_{\rm circ} = 148 \pm 5$ km~s$^{-1}$ from the stellar LOS velocity and velocity dispersion maps using the asymmetric drift equation \citep{Binney1987} as was done in \citet{Debattista2002}. We assumed the radial, azimuthal, and vertical components of the velocity dispersion having exponential radial profiles with the same scalelength, but different central values and following the epicyclic approximation. We also adopted a constant circular velocity. For our dynamical model, we selected all the spatial bins within the elliptical annulus mapping the inner disc and characterised by $a_{\rm min}=13''$ (corresponding to the projection of the bar length along the disc major axis), $a_{\rm max}=36''$ (corresponding to the disc break radius), and $\epsilon = 0.242$. We adopted the scalelength of the inner disc ($h_{\rm in}=11\farcs8\pm0\farcs1$) from the photometric decomposition. The comparison between the observed and modelled kinematic maps to derive $V_{\rm circ}$ is shown in Fig.~\ref{fig:kinematics}.

\subsection{Bar pattern speed}
\label{sec:tw}

We applied the TW method to the sky-cleaned MUSE datacube of NGC\,4277 to measure its bar pattern speed. We defined nine adjacent pseudo-slits crossing the bar and aligned with the disc. They have a width of nine pixels ($1\farcs8$) to account for seeing smearing effects and a half length of 175 pixels ($35''$) to cover the extension of the inner disc, and ${\rm PA}=123\fdg27$. 

We derived the photometric integrals $\langle X\rangle$ from the MUSE reconstructed image, which we obtained by summing the MUSE datacube along the spectral direction in the same wavelength range adopted to measure the stellar kinematics. In each pseudo-slit, we measured the luminosity-weighted position of the stars with respect to the galaxy minor axis as follows:

\begin{equation}
  \langle X\rangle = \frac{\sum_{(x,y)} F(x,y)\,
  {\rm dist}(x,y)}{\sum_{(x,y)} F(x,y)},
  \nonumber
\end{equation}
where $(x,y)$ and $F(x,y)$ are the sky-plane coordinates and flux of the pixels in the pseudo-slit, respectively, and dist$(x,y)$ is the distance of the pixels to the pseudo-slit centre (Fig.~\ref{fig:tw}, left panel). We adopted the $i$-band SDSS Petrosian radius as the galaxy effective radius $R_{\rm e}=13\farcs8$ and we checked the convergence of the $\langle X \rangle$ integrals by measuring their values for different pseudo-slit lengths ranging from $1.3 R_{\rm e} = 17\farcs9$ to $35''$ (Fig.~\ref{fig:stability}, left panel). In this radial range, $\langle X \rangle$ are expected to be constant \citep{Zou2019}; we adopted their root mean square as the $1\sigma$ error on $\langle X \rangle$.

We derived the kinematic integrals $\langle V\rangle$ subtracted of the galaxy systemic velocity in the same wavelength range adopted for the stellar kinematics. We summed all the spaxels of each pseudo-slit to obtain a single spectrum from which we measured the luminosity-weighted stellar LOS velocity with {\sc ppxf}. This is equivalent to calculating the following:

\begin{equation}
  \langle V\rangle = \frac{\sum_{(x,y)}V_{\rm LOS}(x,y)\, F(x,y)}{\sum_{(x,y)} F(x,y)},
\nonumber
\end{equation} 
where $(x,y)$ and $V_{\rm LOS}(x,y)$ are the coordinate of the spaxels in the pseudo-slit and their stellar LOS velocity, respectively (Fig.~\ref{fig:tw}, central panel). We adopted the rescaled formal errors by {\sc ppxf} as a $1\sigma$ error on $\langle V \rangle$. We checked the convergence of $\langle V\rangle$ integrals by measuring their values as a function of the pseudo-slit length as was done for the photometric integrals (Fig.~\ref{fig:stability}, right panel). 

Using the {\sc fitexy} routine in {\sc idl}, we fitted the $\langle X \rangle$ and $\langle V \rangle$ integrals with a straight line with a slope $\Omega_{\rm bar} \sin{i} = 2.65\pm 0.37$ km~s$^{-1}$~arcsec$^{-1}$ (Fig.~\ref{fig:tw}, right panel). From the galaxy inclination, we obtained $\Omega_{\rm bar} = 4.06\pm0.56$ km~s$^{-1}$~arcsec$^{-1}$, which corresponds to $\Omega_{\rm bar} = 24.7\pm3.4$ km~s$^{-1}$~kpc$^{-1}$. We calculated the co-rotation radius from the bar pattern speed and circular velocity as $R_{\rm cor} = V_{\rm circ}/\Omega_{\rm bar} = 36\farcs5 \pm 5\farcs2$, which corresponds to $R_{\rm cor} = 6.0 \pm 0.9$ kpc with the $1\sigma$ error estimated from the propagation of uncertainty. Finally, we derived the rotation rate ${\cal{R}}=R_{\rm cor}/a_{\rm bar} = 1.8^{+0.5}_ {-0.3}$ with the $\pm1\sigma$ error estimated from a Monte Carlo simulation to account for the errors on $a_{\rm bar}$, $\Omega_{\rm bar}$, and $V_{\rm circ}$. We conclude that NGC\,4277 hosts a slow bar and this result does not depend on the distance of the galaxy.

\section{Conclusions}
\label{sec:conclusions}

We measured the broad-band surface photometry and two-dimensional stellar kinematics of NGC\,4277, a barred lenticular galaxy at 33.9 Mpc in the region of the Virgo cluster, to derive the pattern speed of its bar ($\Omega_{\rm bar}=24.7\pm3.4$ km s$^{-1}$ kpc$^{-1}$) and the ratio of the co-rotation radius to the bar length (${\cal{R}}=1.8^{+0.5}_ {-0.3}$). NGC\,4277 hosts a weak ($S_{\rm bar}=0.21 \pm 0.02$) and short bar ($R_{\rm bar}=3.2^{+0.9}_{-0.6}$\,kpc), which falls short of the co-rotation radius ($R_{\rm cor} = 6.0 \pm 0.9$ kpc). 
This is a remarkable result and we carefully handled the sources of error affecting the measurements of the TW integrals by combining the deep SDSS imaging to the wide-field and fine spatially sampled MUSE spectroscopy. As a consequence, the values of $\Omega_{\rm bar}$ and ${\cal{R}}$ of NGC\,4277 are amongst the best-constrained ones ever obtained with the TW method with relative statistical errors of $\sim0.2$. These results hold even if we adopt the galaxy inclination for a thick (with an axial ratio $q_0=0.3$, \citealt{Mosenkov2015}) rather than infinitesimally-thin stellar disc. Indeed, the systematic difference between the inclination-dependent parameters is much smaller than their statistical errors.

\begin{figure}
    \centering
    \includegraphics[scale=0.295]{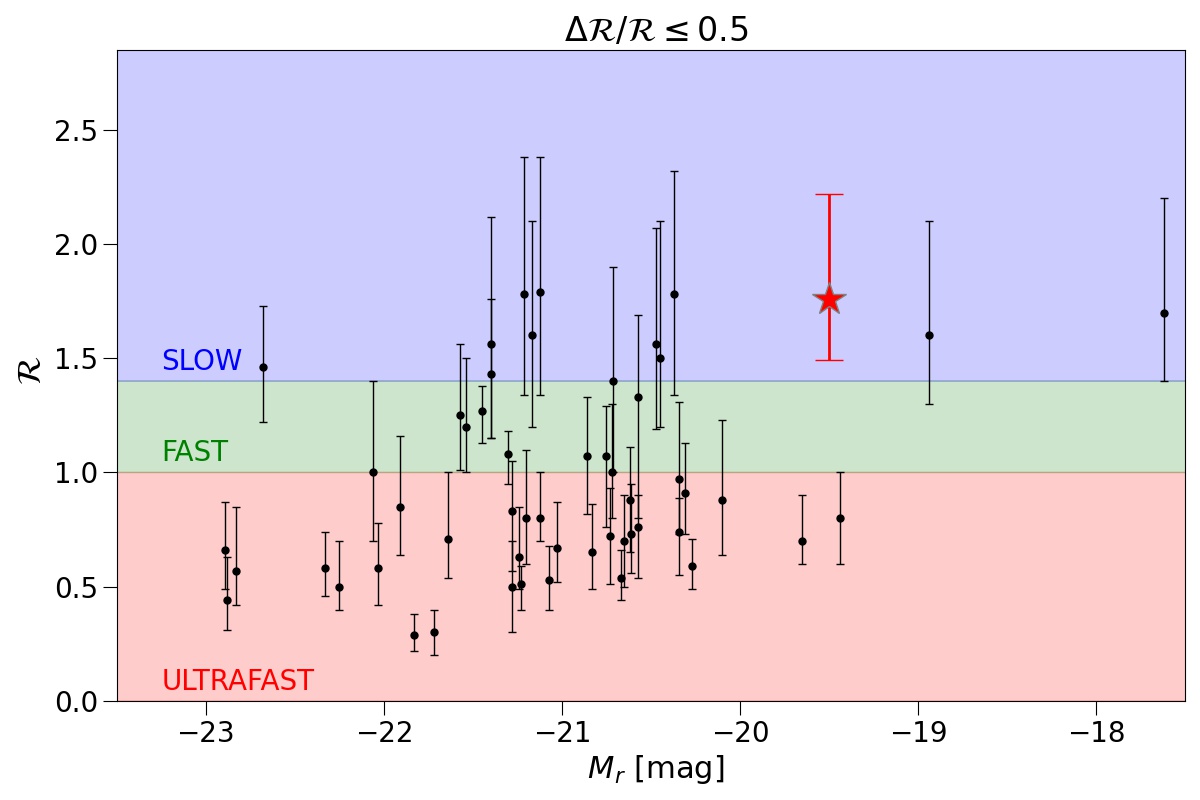}
    \caption{Bar rotation rate as a function of the total $r$-band absolute magnitude for barred galaxies for which the bar pattern speed was measured with the TW method. Only galaxies with $\Delta\mathcal{R}/\mathcal{R}\leq0.5$ are shown \citep{Cuomo2020}. The red star corresponds to NGC\,4277. The coloured regions highlight the ultra-fast (red), fast (green), and slow bar (blue) regimes, respectively.}
    \label{fig:comparison}
\end{figure}

We show in Fig.~\ref{fig:comparison} all the galaxies for which $\Omega_{\rm bar}$ has been measured with the TW method and with a well-constrained rotation rate ($\Delta{\cal{R}}/{\cal{R}}<0.5$, \citealt{Cuomo2020}). Most bars are consistent with being fast within errors ($1\leq \mathcal{R} \leq 1.4$), including the dwarf lenticular IC\,3167 ($M_r=-17.62$ mag and ${\cal{R}}=1.7^{+0.5}_{-0.3}$, \citealt{Cuomo2022lopsided}), whose lopsided bar is twice more likely to be slow (probability of $68\%$) rather than fast ($32\%$). For comparison, the probability of the bar of NGC\,4277 to be slow ($91\%$) is ten times higher than that of being fast ($9\%$). On the other hand, there are many ultra-fast bars ($\mathcal{R}<1$), although this is a non-physical result for a self-consistent bar because the stellar orbits beyond $R_{\rm cor}$ are not aligned with the bar and cannot support it. Recently, \cite{Cuomo2021ultrafast} reanalysed a sub-sample of ultra-fast bars and conclude that their ${\cal{R}}$ was underestimated because of an overestimate of $R_{\rm bar}$.

The only other galaxy nominally hosting a stellar slow bar was manga 8317-12704 ($\mathcal{R}=2.4^{+0.8}_{-0.6}$) and measured by \citet{Guo2019}, who applied the TW method to the stellar kinematics of a sample of barred galaxies from the MANGA survey \citep{MANGA}. However, they adopted a slit semi-length equal to $1.2 R_{\rm eff}$, which does not guarantee the convergence of TW integrals when the bar length is longer than the galaxy effective radius. \citet{Garma-Oehmichen2020} show that $\Omega_{\rm bar}$ of manga 8317-12704 was underestimated (and thus $\mathcal{R}$ was overestimated) because $R_{\rm bar} = 10\farcs3 > 1.2 R_{\rm eff} = 8\farcs6$. They adopted a different PA ($\Delta$PA$\sim3^\circ$) and larger semi-length for the pseudo-slits to obtain stable TW integrals from the MANGA dataset. \citet{Garma-Oehmichen2020} found a new rotation rate for the bar of manga 8317-12704 ($\mathcal{R}=1.5^{+0.3}_{-0.2}$), which is consistent with the fast regime.
The slow bars of NGC\,2915 ($\mathcal{R}=1.7$, \citealt{Bureau1999}), UGC\,628 ($\mathcal{R}=2.0^{+0.5}_{-0.3}$, \citealt{Chemin2009}), and DDO\,168 ($\mathcal{R}=2.1$, \citealt{Patra2019}) cannot be safely taken into account since a gaseous tracer might not satisfy the continuity equation linking the TW integrals.

We conclude that NGC\,4277 is the first clear case of a galaxy hosting a slow stellar bar (${\cal{R}}>1.4$ at the $1.3\sigma$ confidence level) measured with the TW method. By determining the DM fraction in the bar region, it will be possible to understand whether the uncommonly large ${\cal{R}}$ of NGC\,4277 was initially imprinted  by a tidal interaction with NGC\,4273 triggering the bar formation \citep{Martinez-Valpuesta2017} or whether it is the end result of the bar braking due to the dynamical friction exerted by the DM halo \citep{Weinberg1985, Debattista2000, Athanassoula2003, Algorry2017}. 

\begin{acknowledgements}
We thank the anonymous referee, whose helpful comments helped us to improve this manuscript. CB acknowledges the Jeremiah Horrocks Institute for hospitality while this Letter was in progress. CB, EMC, EDB, and AP are supported by MIUR grant PRIN 2017 20173ML3WW-001 and Padua University grants DOR2019-2021. VC is supported by Fondecyt Postdoctoral programme 2022 and by ESO-Chile Joint Committee programme ORP060/19. JALA and LC are supported by the Spanish Ministerio de Ciencia e Innovaci\'on y Universidades by the grants PID2020-119342GB-I00 and PGC2018-093499-B-I00, respectively. JMA acknowledges the support of the Viera y Clavijo Senior program funded by ACIISI and ULL.
\end{acknowledgements}

\bibliographystyle{aa} 
\bibliography{biblio} 

\begin{appendix}

\section{Isophotal analysis}
\label{app:ellipse}

\renewcommand{\thefigure}{A\arabic{figure}}
\setcounter{figure}{0}

\begin{figure*}
    \centering
    \includegraphics[scale=0.36]{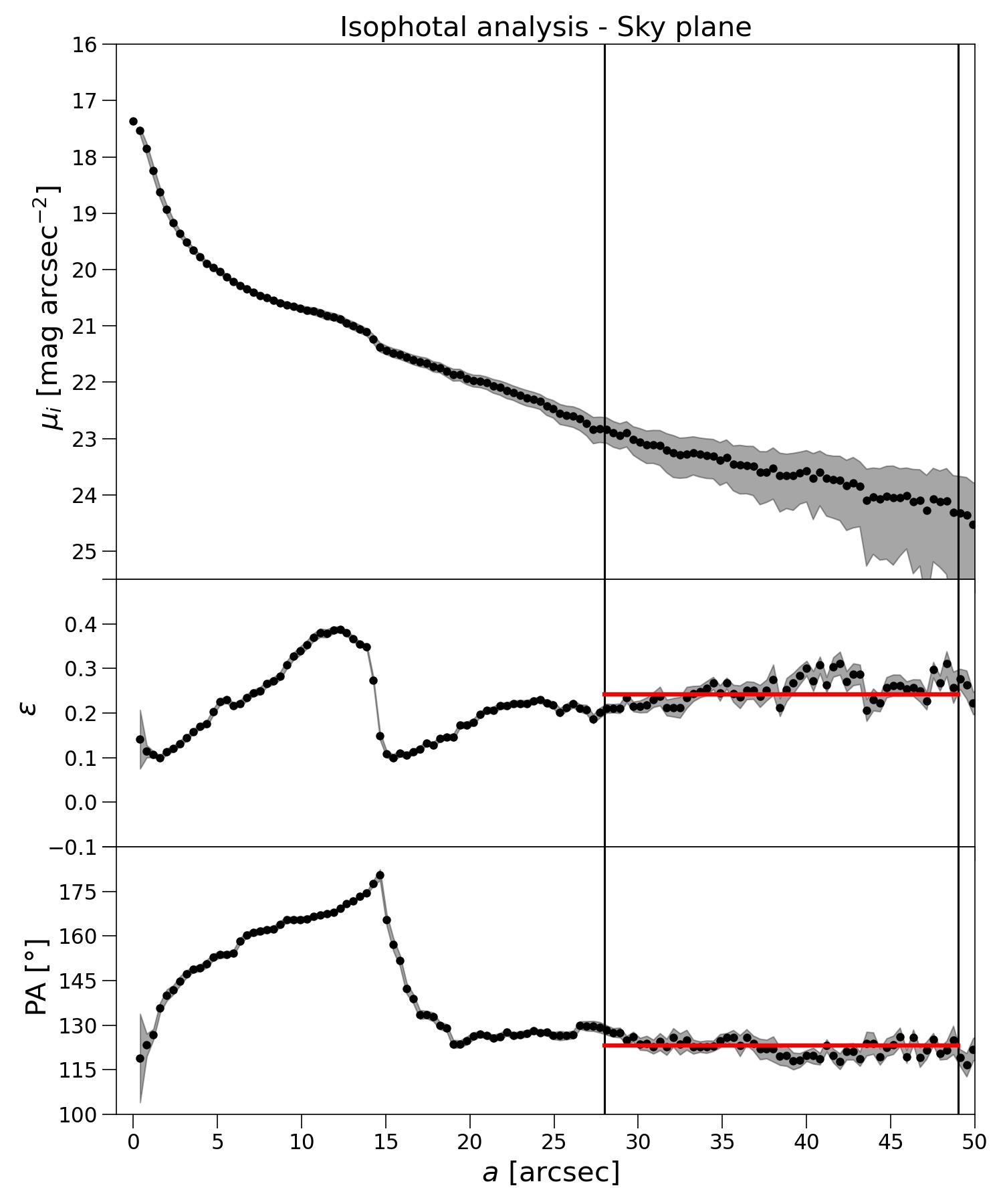}
    \includegraphics[scale=0.36]{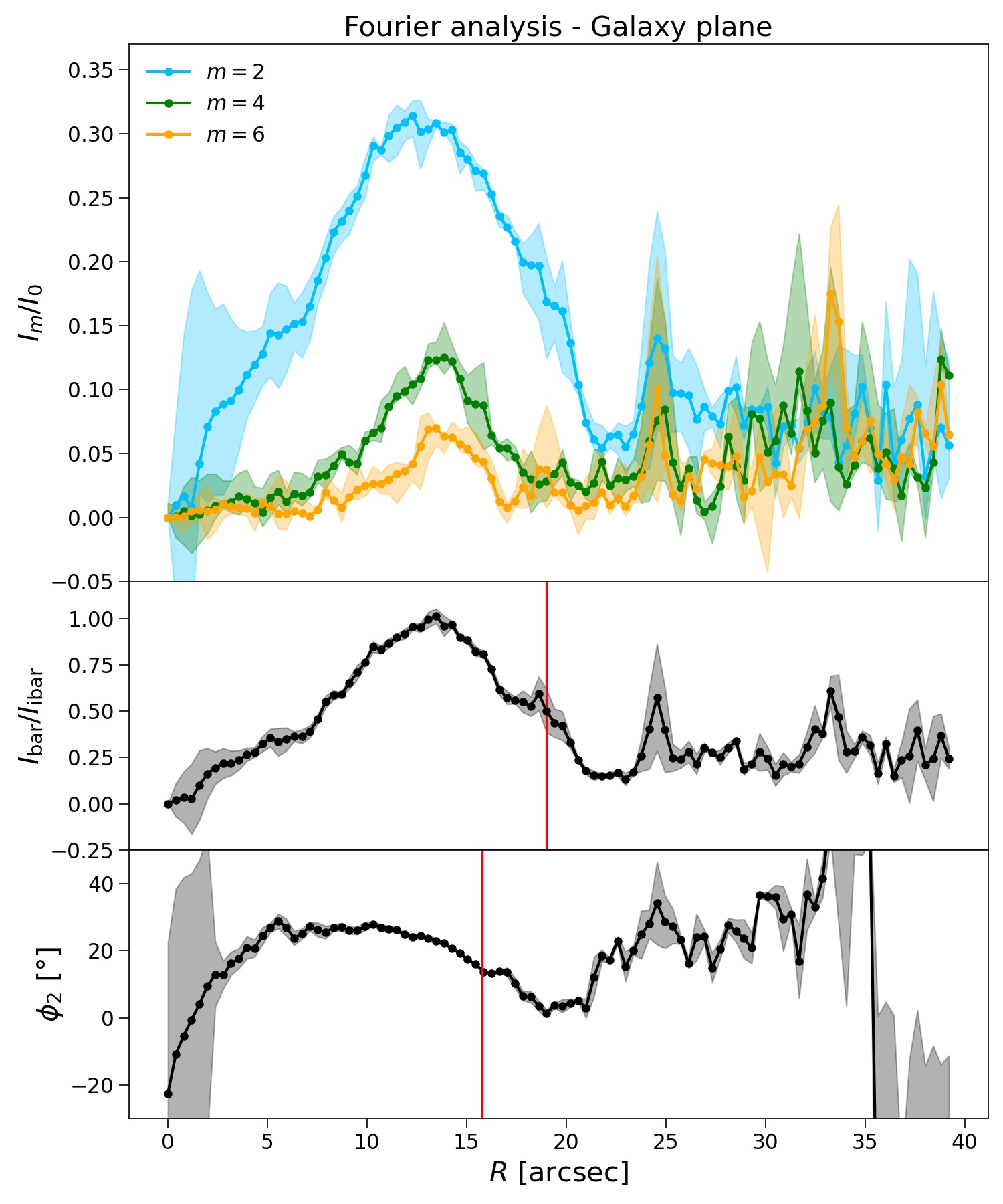}
    \caption{{\itshape Left panels}: Isophotal analysis of the $i$-band image of NGC\,4277. The radial profiles of the surface brightness ({\em upper panel\/}), position angle ({\em central panel\/}), and ellipticity ({\em lower panel\/}) are shown as a function of the semi-major axis of the best-fitting isophotal ellipses. The vertical black lines bracket the radial range adopted to estimate the mean ellipticity ($\langle\epsilon\rangle$ = 0.242 $\pm$ 0.002) and position angle ($\langle$PA$\rangle$ = 123\fdg27 $\pm$ 0\fdg32) of the disc. {\itshape Right panels}: Fourier analysis of the deprojected $i$-band image of NGC\,4277. The radial profiles of the relative amplitude of the $m=2$ (blue points), $m=4$ (green points), and $m=6$ (yellow points) Fourier components ({\em upper panel\/}), bar-interbar intensity ratio ({\em central panel\/}), and phase angle $\phi_2$ of the $m=2$ Fourier component ({\em lower panel\/}) are shown as a function of galactocentric distance. The vertical red lines in the central and lower panels mark the bar radii $R_{\rm bar/ibar}$ and $R_{\phi_2}$, respectively.}
    \label{fig:photometry}
\end{figure*}

\renewcommand{\thefigure}{C\arabic{figure}}
\setcounter{figure}{0}

\begin{figure*}
    \centering
    \includegraphics[scale=0.28]{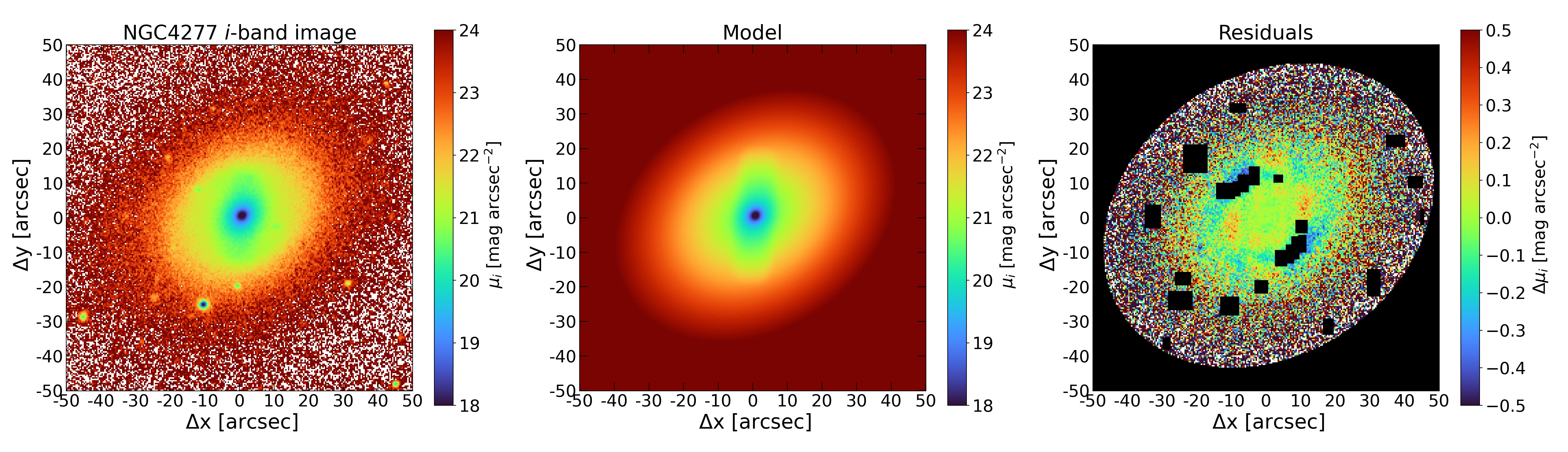}
    \caption{Photometric decomposition of the $i$-band image of NGC\,4277 with the maps of the observed ({\em left panel\/}), model ({\em central panel\/}), and residual (observed$-$model) surface-brightness distribution ({\em right panel\/}). The FOV of the images is oriented with north being up and east to the left.}
    \label{fig:gasp2d}
\end{figure*}

We analysed the flux-calibrated and sky-subtracted $i$-band image of NGC\,4277. The choice of $i$-band ensured that we reached a sufficient spatial resolution ($\rm FWHM=1\farcs4$) and depth ($\mu_{i,{\rm sky}} = 20.46\pm0.04$ mag arcsec$^{-2}$), and minimised the dust effects with respect to the other SDSS passbands to characterise the bar component with an accurate photometric decomposition of the surface-brightness distribution.

We masked all the foreground stars, background galaxies, and spurious sources (such as residual cosmic rays and bad pixels) in the image FOV and fitted the galaxy isophotes using the {\sc ellipse} task in {\sc iraf} \citep{Jedrzejewski1987}. First, we allowed the centre, ellipticity, and position angle of the fitting ellipses to vary. Then, we fitted the isophotes again but with {\sc ellipse}, adopting the centre of the inner ellipses. The resulting radial profiles of the azimuthally averaged surface brightness, $\mu_i$, position angle, PA, and ellipticity, $\epsilon$, are shown in Fig.~\ref{fig:photometry} (left panels). We did not correct the measured surface brightness for cosmological dimming ($z=0.00730$, NED), Galactic absorption ($A_i=0.032$ mag, \citealt{Schlafly2011}), or K correction ($K_i=0.01$ mag, \citealt{Chilingarian2010}).

We derived the mean values of $\epsilon$ and PA of the disc in the radial range $28''\leq a \leq 48''$ (Fig.~\ref{fig:photometry}, left panels), which extends outside the bar-dominated region to the farthest fitted isophote. We defined the extension of this radial range by fitting the PA measurements with a straight line and considering all the radii where the line slope was consistent with being zero within the associated root mean square error, as was done by \citet{Cuomo2019}. 

We obtained the bar length from the isophotal analysis of the deprojected $i$-band image of NGC\,4277, which we built by stretching the image along the disc minor axis ($\rm PA=33\fdg27$) by a factor $1/\cos{i}$ where $i$ is the disc inclination. As in \citet{Aguerri2009}, we defined the bar length $R_{\rm PA}$ as the radius at which we measured $\rm \Delta PA=10\degr$ with respect to the PA of the ellipse with the maximum $\epsilon$.
We also calculated the bar strength following \citet{Aguerri2009}:

\begin{equation}
\nonumber
\displaystyle S_\epsilon = \frac{2}{\pi} \left[ \arctan{(1 - \epsilon_\textrm{bar})^{-1/2}} - \arctan{ (1 - \epsilon_\textrm{bar})^{1/2}} \right],
\end{equation}

\noindent
where $\epsilon_{\rm bar}$ is the bar ellipticity measured at $R_{\rm bar}$ as obtained in Sect.~\ref{sec:analysis}. We estimated the error with a Monte Carlo simulation by accounting for the error on the ellipticity. We took the standard deviation of the resulting distribution as the statistical error on $S_\epsilon$.

\section{Fourier analysis}
\label{app:fourier}

We performed the Fourier analysis of the deprojected $i$-band image of NGC\,4277 and decomposed its azimuthal surface-brightness distribution as  

\begin{equation}
I(R,\phi)=\frac{A_0(R)}{2}+\sum_{m=1}^{\infty}[A_m(R)\cos{(m\phi)}+B_m(R)\sin{(m\phi)}],
\nonumber
\end{equation}

\noindent where $R$ is the galactocentric radius on the galaxy plane and $\phi$ is the azimuthal angle measured anticlockwise from the line of nodes, while the Fourier coefficients are as follows:

\begin{eqnarray}
A_m(R)=\frac{1}{\pi}\int_0^{2\pi} I(R,\phi)\cos{(m\phi)}\,{\rm d}\phi \nonumber \\ B_m(R)=\frac{1}{\pi}\int_0^{2\pi} I(R,\phi)\sin{(m\phi)}\,{\rm d}\phi.
\nonumber
\end{eqnarray}

\noindent We obtained the radial profiles of the amplitudes of the $m=0,2,4,6$ Fourier components as follows:

\begin{equation}
I_m(R)= \begin{cases} 
A_0(R)/2  & \mbox{if }m = 0 \\ 
\sqrt{A_m^2(R)+B_m^2(R)} & \mbox{if }m \neq 0.
\nonumber
\end{cases}
\end{equation}

\noindent We then derived the radial profile of the intensity contrast between the bar and interbar regions and defined the bar length $R_{\rm bar/ibar}$ as the largest radius where

\begin{equation}
\nonumber
\frac{ I_{\rm bar}}{I_{\rm ibar}} = \frac{I_0+I_2+I_4+I_6}{I_0-I_2+I_4-I_6}>\frac{1}{2} \left[\max\left( \frac{I_{\rm bar}}{I_{\rm ibar}} \right)+\min \left( \frac{I_{\rm bar}}{I_{\rm ibar}}\right) \right].
\end{equation}

\noindent The radial profiles of the relative amplitudes of the $m=2, 4, 6$ Fourier components, phase angle $\phi_2$ of the $m=2$ Fourier component, and bar and interbar intensity are shown in Fig.~\ref{fig:photometry} (right panels).

Finally, we calculated the bar strength as the mean value of the $I_2/I_0$ ratio over the bar extension as follows:

\begin{equation}
\nonumber
S_{\textrm{Fourier}} =\frac{1}{R_{\textrm{bar}}} \int^{R_{\textrm{bar}}}_0 \frac{I_2}{I_0}(R)\,{\rm d}R,
\end{equation}

\noindent as was done in \citet{Athanassoula2002} and adopting $R_{\rm bar}$ from Sect.~\ref{sec:analysis}. We estimated the error by performing a Monte Carlo simulation and taking the errors on the $m=0,2$ Fourier components into account. We generated 100 mock profiles of the $I_2/I_0$ intensity ratio and we calculated the corresponding bar strength. We took the standard deviation of the resulting distribution as the statistical error on $S_{\rm Fourier}$.

\renewcommand{\thefigure}{D\arabic{figure}}
\setcounter{figure}{0}

\begin{figure*}
    \centering
    \includegraphics[scale=0.4]{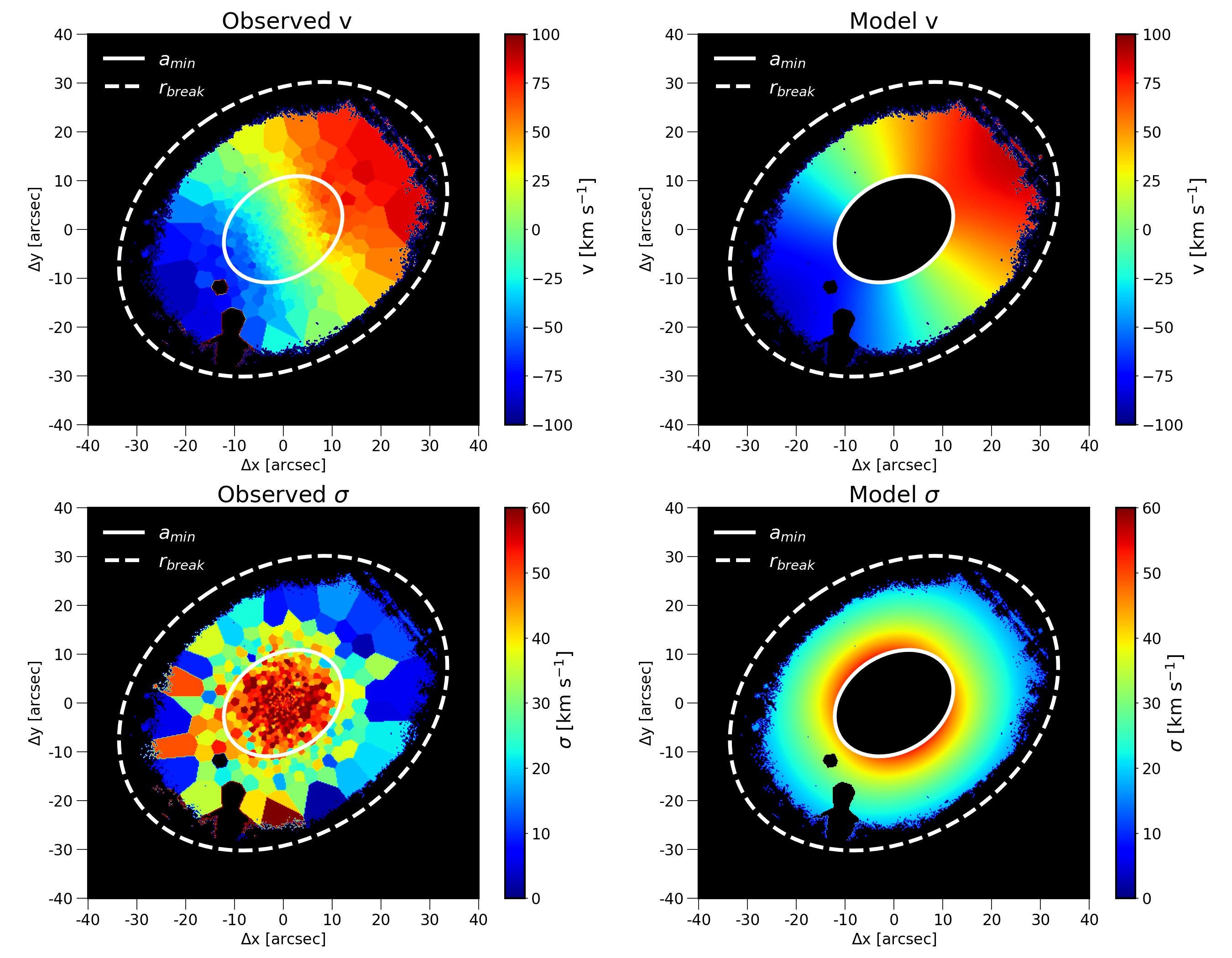}
    \caption{Maps of the stellar LOS velocity subtracted of systemic velocity ({\em top panels\/}) and velocity dispersion corrected for $\sigma_{\rm inst}$ ({\em bottom panels\/}) of NGC~4277 derived from the $S/N = 40$ Voronoi-binned MUSE data ({\em left panels\/}) and from the asymmetric-drift-corrected dynamical model ({\em right panels\/}). The FOV is 1\farcm3\,$\times$\,1\farcm3 and is oriented with north being up and east to the left. The solid and dashed white lines mark the region adopted for modelling between the inner edge of the inner disc and location of the disc break radius, respectively.}
    \label{fig:kinematics}
    
\end{figure*}

\section{Photometric decomposition}
\label{app:gasp2d}

\begin{table}[!ht]
\centering
\footnotesize
\renewcommand{\arraystretch}{1.2}
\caption{Structural parameters of NGC\,4277 from the photometric decomposition. The scalelengths are not deprojected on the galactic plane.}
\begin{tabular}{cc}
\hline
\multicolumn{2}{ c }{Bulge}\\
\hline
$\mu_\textrm{e}$ & 19.29 $\pm$ 0.03 mag arcsec$^{-2}$ \\
$r_\textrm{e} $  & 1.75 $\pm$ 0.03 arcsec \\
$n$ & 2.36 $\pm$ 0.03 \\
$q_\textrm{bulge}$ & 0.841 $\pm$ 0.004 \\
PA$_\textrm{bulge}$ & 135\fdg09 $\pm$ 0\fdg05 \\
$L_\textrm{bulge}/L_\textrm{T}$ & 0.11 \\
 \hline
\multicolumn{2}{ c }{Disc}\\
\hline
$\mu_0$ & 20.14 $\pm$ 0.01 mag arcsec$^{-2}$  \\
$h_\textrm{in}$ & 11.82 $\pm$ 0.10 arcsec \\
$h_\textrm{out}$ & 14.81 $\pm$ 0.22 arcsec  \\
$r_\textrm{break}$ & 36.16 $\pm$ 0.57 arcsec \\
$q_\textrm{disc}$ & 0.758 $\pm$ 0.001 \\
PA$_{\textrm{disc}}$ & 123\fdg37 $\pm$ 0\fdg06\\
$L_\textrm{disc}/L_\textrm{T}$ & 0.82 \\
\hline
\multicolumn{2}{ c }{Bar}\\
\hline
$\mu_0$ & 21.37 $\pm$ 0.01 mag arcsec$^{-2}$ \\
$a_\textrm{bar}$ & 20.70 $\pm$ 0.07 arcsec\\
$q_\textrm{bar}$ & 0.341 $\pm$ 0.001\\
PA$_{\rm bar}$ & 175\fdg59 $\pm$ 0\fdg04 \\
$L_\textrm{bar}/L_\textrm{T}$ & 0.07\\
\hline
\label{tab:gasp2d}
\end{tabular}
\end{table}

We derived the structural parameters of NGC\,4277 by applying the {\sc gasp2d} algorithm \citep{Mendez-Abreu2008, MendezAbreu2017, MendezAbreu2018} to the flux-calibrated and sky-subtracted $i$-band image of the galaxy. We modelled the galaxy surface brightness in each pixel of the image to be the sum of the light contribution of a S\'ersic bulge, a double-exponential disc, and a Ferrers bar. We did not account for other luminous components, such as rings or spiral arms. We assumed that the isophotes of each component are elliptical and centred on the galaxy centre with constant values for the position angle and axial ratio.
We parameterised the bulge surface brightness as

\begin{equation}
I_{\rm bulge}(x,y)=I_{\rm e}10^{-b_n[(r_{\rm bulge}/r_{\rm e})^{1/n}-1]},
\nonumber
\end{equation}

\noindent following \citet{Sersic1968}, where $(x,y)$ are the Cartesian coordinates of the image in pixels, $r_{\rm e}$ is the effective radius encompassing half of the bulge light, $I_{\rm e}$ is the surface brightness at $r_{\rm e}$, $n$ is the shape parameter of the surface brightness profile, and $b_n=0.868n-0.142$ is a normalisation
coefficient \citep{caon1993}. The radius $r_{\rm bulge}$ is defined as follows:

\begin{equation}
\begin{aligned}
\nonumber
r_\textrm{bulge}= [(-(x-x_0) \sin{\mathrm{PA_{bulge}}} + (y-y_0) \cos{\mathrm{PA_{bulge}}})^2 \\ + \ ((x-x_0)\cos {\mathrm{PA_{bulge}}} + (y-y_0)\sin{\mathrm{PA_{bulge}}})^2 /q^2_\textrm{bulge}]^{1/2},
\end{aligned}
\end{equation}

\noindent where $(x_0, y_0)$, PA$_{\rm bulge}$, and $q_{\rm bulge}$ are the coordinates of the galaxy centre, bulge position angle, and bulge axial ratio, respectively. 
We parameterised the disc surface brightness as 

\begin{equation}
I_{\rm disc}(x,y)=
\begin{cases}   I_{\rm 0}e^{-r_{\rm disc}/h_{\rm in}} , & \mbox{if } r \leq r_{\rm break} \\ I_{\rm 0} e^{-r_{\rm break}(h_{\rm out}-h_{\rm in})/h_{\rm out}} e^{-r/h_{\rm out}}  & \mbox{if }r > r_{\rm break},
\nonumber
\end{cases}
\end{equation}

\noindent following \citet{MendezAbreu2017}, where $I_{\rm 0}$ is the central surface brightness, $r_{\rm break}$ is the break radius at which the surface brightness profile changes slope, and $h_{\rm in}$ and $h_{\rm out}$ are the scalelengths of the inner and outer exponential profile, respectively. The radius $r_{\rm disc}$ is defined as follows:

\begin{equation}
\begin{aligned}
\nonumber
r_\textrm{disc}= [(-(x-x_0) \sin{\mathrm{PA_\textrm{disc}}} + (y-y_0) \cos{\mathrm{PA_\textrm{disc}}})^2 \\ + \ ((x-x_0)\cos {\mathrm{PA_\textrm{disc}}} + (y-y_0)\sin{\mathrm{PA_\textrm{disc}}})^2 /q^2_\textrm{disc}]^{1/2},
\end{aligned}
\end{equation}

\noindent where PA$_{\rm disc}$ and $q_{\rm disc}$ are the disc position angle and axial ratio, respectively. 
We parameterised the bar surface brightness as follows:

\begin{equation}
I_{\rm bar}(r)=
\begin{cases} 
I_{\rm 0,bar}\left[1-(r_{\rm bar}/a_{\rm bar})^2 \right]^{2.5} & \mbox{if }r_{\rm bar} \leq a_{\rm bar} \\ 0 & \mbox{if }r_{\rm bar} > a_{\rm bar},
\nonumber
\end{cases}
\end{equation}

\noindent following \citet{Aguerri2009}, where $I_{\rm 0,bar}$ and $a_{\rm bar}$ are the bar central surface brightness and length, respectively. The radius $r_{\rm bar}$ is defined as 

\begin{equation}
\begin{aligned}
\nonumber
r_\textrm{bar}= [(-(x-x_0) \sin{\mathrm{PA_{bar}}} + (y-y_0) \cos{\mathrm{PA_{bar}}})^2 \\ + \ ((x-x_0)\cos {\mathrm{PA_{bar}}} + (y-y_0)\sin{\mathrm{PA_{bar}}})^2 /q^2_\textrm{bar}]^{1/2},
\end{aligned}
\end{equation}

\noindent where PA$_{\rm bar}$ and $q_{\rm bar}$ are the bar position angle and axial ratio, respectively. The best-fitting values of the structural parameters of the bulge, disc, and bar are returned by {\sc gasp2d} by performing a $\chi^2$ minimisation. Figure~\ref{fig:gasp2d} shows the $i$-band image, {\sc gasp2d} best-fitting image, and residual image of NGC\,4277. We estimated the errors on the best-fitting structural parameters by analysing the images of a sample of mock galaxies generated by \citet{MendezAbreu2017} with Monte Carlo simulations and mimicking the instrumental setup of the available SDSS image (but see also \citealt{Costantin2017}). 

We adopted the mean and standard deviation of the relative errors of the mock galaxies as the systematic and statistical errors of the parameters of the surface-brightness radial profiles of the bulge ($I_{\rm e}$, $r_{\rm e}$, and $n$), disc ($I_{0,{\rm disc}}$, $h_{\rm in}$, $h_{\rm out}$, and $r_{\rm break}$), and bar ($I_{0,{\rm bar}}$ and $a_{\rm bar}$). We adopted the mean and standard deviation of the absolute errors of the mock galaxies as the systematic $\sigma_{\rm syst}$ and statistical $\sigma_{\rm stat}$ errors of the geometric parameters of the bulge (PA$_{\rm bulge}$ and $q_{\rm bulge}$), disc (PA$_{\rm disc}$ and $q_{\rm disc}$), and bar (PA$_{\rm bar}$ and $q_{\rm bar}$). We computed the errors on the best-fitting parameters as $\sigma^2 = \sigma^2_{\rm stat} + \sigma^2_{\rm syst}$, with the systematic errors being negligible compared to the statistical ones. The quoted uncertainties are purely formal and do not take into account the parameter degeneracy and a different parameterisation of the components. The values of the best-fitting structural parameters of NGC\,4277 and corresponding errors are reported in Table~\ref{tab:gasp2d}.

\section{Dynamical analysis}
\label{app:dynamics}

We derived the circular velocity $V_{\rm circ}$ from the stellar LOS velocity and velocity dispersion in the region of the inner disc using the asymmetric drift equation \citep{Binney1987}. We selected the spatial bins within an elliptical annulus with semi-major axes $a_{\rm min} = 13''$ and $a_{\rm max} = r_{\rm break} = 36''$ and ellipticity $\epsilon = 0.242$ (Fig.~\ref{fig:kinematics}) and followed the prescriptions of \citet{Debattista2002} and \citet{Aguerri2003} to obtain the following:

\begin{equation}
v(r,\theta)=\sqrt{V_{\rm circ}^2(R)+\sigma_{R}^2(R)\left[1- \frac{\sigma_\phi^2(R)}{\sigma_R^2(R)}-R\left(\frac{1}{h}+\frac{2}{a}\right)\right]}\cos\phi\sin{i}
\nonumber
\end{equation}

\begin{equation}
\nonumber \sigma(r,\theta)=\sigma_{R}(R)\sqrt{ \sin^2{i}\left[\sin^2\phi+\frac{\sigma_\phi^2(R)}{\sigma_R^2(R)}
\cos^2\phi\right]+\frac{\sigma^2_{0,z}}{\sigma^2_{0,R}}\cos^2{i} },
\end{equation}

\noindent where $r$ is the galactocentric radius on the sky plane and $\theta$ is the anomaly angle measured anticlockwise from the line of nodes. The polar coordinates defined on the galaxy ($R$, $\phi$) and sky plane ($r$, $\theta$) are related to each other as follows:
\begin{equation}
R \cos \phi =r \cos \theta \quad , \quad 
\tan \phi \cos i =\tan \theta.
\nonumber
\end{equation}
\noindent We adopted $h = h_{\rm in}\,=\,11\farcs82$ and
$i\,=\,40\fdg7$ and we assumed the three components of the velocity dispersion to have exponential radial profiles with the same scalelength, but different central values: 
\begin{equation}
\sigma_{R} = \sigma_{0,R}\, e^{-R/a} \quad , \quad 
\sigma_{\phi} = \sigma_{0,\phi}\, e^{-R/a}  \quad , \quad 
\sigma_{z} = \sigma_{0,z}\,e^{-R/a}. 
\nonumber
\end{equation}
\noindent This means that the shape of the velocity ellipsoid does not change with the galactocentric radius having constant axial ratios 
$(\sigma_\phi/\sigma_R, \sigma_z/\sigma_R) = (\sigma_{0,\phi}/\sigma_{0,R}, \sigma_{0,z}/\sigma_{0,R})$. Then, we parameterised the circular velocity with the following power law: 
\begin{equation}
V_{\rm circ} = V_0\,R^\alpha. 
\nonumber
\end{equation}

\noindent Assuming the epicyclic approximation ($\sigma_\phi/\sigma_R = \sqrt{0.5(1+\alpha)}$) and a constant circular velocity ($\alpha=0$), we found $V_{\rm circ} = 148 \pm 5$ km~s$^{-1}$. The maps of the disc stellar kinematics with the best-fitting LOS velocity and velocity dispersion are shown in Fig.~\ref{fig:kinematics}. We need to improve the stellar dynamical modelling to constrain the DM content of NGC\,4277 and get the actual radial profile of its circular velocity. Finding a rising circular velocity will translate into an even larger rotation rate, confirming the main result of this paper.

\section{Tremaine-Weinberg analysis}
\label{app:tw}

\renewcommand{\thefigure}{E\arabic{figure}}
\setcounter{figure}{0}

\begin{figure*}
    \centering
    \includegraphics[scale=0.3]{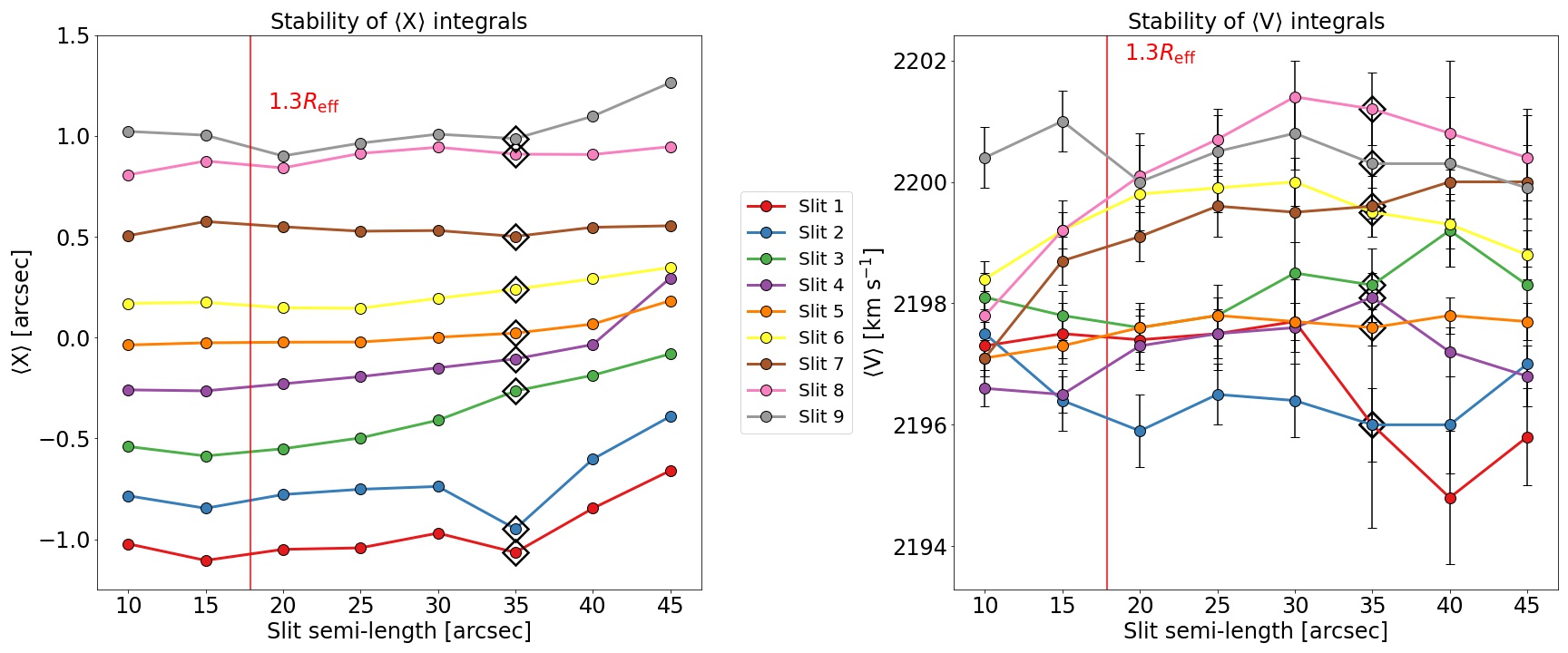}
    \caption{Stability of TW integrals. Photometric ({\em left panel\/}) and kinematic ({\em right panel\/}) integrals as a function of the semi-length of the slit. The adopted values for the TW analysis are marked with empty black diamonds. The vertical red line marks the edge of the region where TW integrals are expected to be constant according to \citet{Zou2019}.}
    \label{fig:stability}
\end{figure*}

\begin{table*}
\centering
\caption{Results of tests on the bar pattern speed and rotation rate of NGC~4277 as a function of the PA of the pseudo-slits.}
\label{tab:test}
\renewcommand{\arraystretch}{1.5}
\begin{center}
\begin{tabular}{c c c c c c}
\hline 
$\sigma_{\langle X \rangle}$ &  $\Omega_{\rm bar}\sin{i}$  & $\Omega_{\rm bar}$ & $\Delta\Omega_{\rm bar}/\Omega_{\rm bar}$ & {$\cal{R}$}  & {$\Delta \cal{R}$/${\cal{R}}$} \\
  &  [km s$^{-1}$ arcsec$^{-1}$] & [km s$^{-1}$ kpc$^{-1}$]    &   &         \\
\hline
\hline
\multicolumn{6}{ c }{PA = 123\fdg27 ($\rm \equiv \langle PA \rangle$) }\\
\hline
MC  & 2.63$\pm$ 0.36  &  24.6$\pm$3.4  & 0.14  & 1.77$^{+0.45}_{-0.27}$  & 0.20 \\ 
rms & 2.65$\pm$ 0.37  &  24.7$\pm$3.4  & 0.14  & 1.76$^{+0.46}_{-0.27}$  & 0.21 \\
\hline
\hline
\multicolumn{6}{ c }{PA = 123\fdg59 ($\rm \equiv \langle PA \rangle + \sigma$)}\\
\hline
rms &  2.35$\pm$ 0.38  &  21.9$\pm$3.6  & 0.16   & 1.88$^{+0.67}_{-0.22}$ & 0.24\\
\hline
\hline
\multicolumn{6}{ c }{PA = 122\fdg95 ($\rm \equiv \langle PA \rangle - \sigma$)}\\
\hline
rms & 2.81$\pm$ 0.38  &  26.3$\pm$3.5   & 0.13   & 1.67$^{+0.41}_{-0.26}$ & 0.20\\
\hline
\hline
\multicolumn{6}{ c }{PA = 121\fdg77 ($\rm \equiv \langle PA \rangle - 1\fdg5$)}\\
\hline
rms & 3.23$\pm$ 0.32 & 30.1$\pm$3.0  & 0.10  & 1.41$^{+0.36}_{-0.16}$  & 0.18 \\
\hline
\hline
\end{tabular}
\end{center}
\end{table*}

We checked the convergence of the photometric integrals as a function of the pseudo-slit semi-length from $10''$ to $45''$ (Fig.~\ref{fig:stability}, left panel). We noticed that the photometric integrals measured for pseudo-slit semi-lengths of $45''$ are systematically larger than those measured at $35''$ and $40''$, which is possibly due to an imperfect subtraction of the sky background at the edges of the FOV of the MUSE datacube. Therefore, we decided to adopt a semi-length of $35''$ for the pseudo-slits crossing the bar. Some of the pseudo-slits cover a few foreground stars, resulting in spurious spikes in the surface-brightness radial profile, which we manually corrected by linearly interpolating over the star light contribution. 
We estimated the errors on $\langle X\rangle$ with a Monte Carlo simulation by generating 100 mock images of the galaxy. To this aim, we processed the convolved, resampled, and reconstructed MUSE image using the {\sc iraf} task {\sc boxcar}. Then, to each image pixel, we added, the photon noise due to the contribution of both sky background and galaxy and the read-out noise of the detector to mimic the actual image of NGC\,4277. We measured the photometric integrals in the mock images and adopted the root mean square of the distribution of measured values as the error for the photometric integral in each pseudo-slit (labelled as `MC' in Table~\ref{tab:test}). As a consistency check, we alternatively estimated the errors on $\langle X\rangle$ defining, for each slit, the radial range in which the value of the photometric integral is constant and adopting the root mean square of the distribution as the error of photometric integrals (labelled as `rms' in Table~\ref{tab:test}). As was done for the photometric integrals, we checked the convergence of the kinematic integrals as a function of the pseudo-slit semi-length from $10''$ to $45''$ (Fig.~\ref{fig:stability}, right panel), and we found that the measured values are compatible within the uncertainties. As kinematic integrals, we chose the values corresponding to the semi-length of $35''$ and we adopted the rescaled formal errors by {\sc ppxf} as associated errors. Our adopted value of $\Omega_{\rm bar}= 24.7\pm3.4$ km s$^{-1}$ kpc$^{-1}$ is consistent with the mean value $\langle \Omega_{\rm bar} \rangle = 21.4\pm1.1$ km s$^{-1}$ kpc$^{-1}$, which we calculated for all the semi-lengths between $20''$ and $35''$ and which corresponds to a slow bar as well.

Even if TW is a model-independent method to recover $\Omega_{\rm bar}$, there are several sources of error which contribute to the resulting accuracy in estimating $\mathcal{R}$ (see \citealt{Corsini2011}, for a discussion). In particular, the misalignment between the orientation of the pseudo-slits and disc PA translates into a large systematic error \citep{Debattista2003}. To account for this issue, we repeated the analysis by adopting different PAs for the pseudo-slits ($\rm \langle PA \rangle - \sigma = 122\fdg95$ and  $\rm \langle PA \rangle + \sigma = 123\fdg59$) to account for the uncertainty on the PA of the inner disc. We obtained the new reconstructed image and defined nine pseudo-slits crossing the bar with a $1\farcs8$ width and a $35''$ semi-length. We manually corrected the surface-brightness radial profile of the pseudo-slit for light contribution of foreground stars, checked the stability of both photometric and kinematic integrals, and derived the bar pattern speed and rotation rate as was done before. The results for the different PAs are listed in Table~\ref{tab:test} and are consistent with a slow bar. As a final test, we repeated the analysis, varying the PA of the pseudo-slits in steps of $\pm0\fdg5$ to look for the PA for which the bar can be classified as fast. This occurs at $\rm \langle PA \rangle - 1\fdg5$ (Table~\ref{tab:test}), which corresponds to a misalignment between the pseudo-slits and disc major axis of $\sim5\sigma$ times the uncertainty on the $\rm \langle PA \rangle$. This is not consistent with the results of the photometric analysis (Fig.~\ref{fig:photometry}) and photometric decomposition (Table~\ref{tab:gasp2d}). All the above consistency checks support the finding of a slow  bar in NGC\,4277.

\end{appendix}

\end{document}